\documentstyle[epsfig,aps,preprint]{revtex}
\tightenlines
\draft
\def\three_j(#1,#2,#3,#4,#5,#6){\pmatrix{#1 & #2 & #3\cr
                                         #4 & #5 & #6\cr}}

\def\qqq{\end{document}}
\def\pmb#1{\setbox0=\hbox{$#1$}%
\kern-.025em\copy0\kern-\wd0
\kern.05em\copy0\kern-\wd0
\kern-.025em\raise.0433em\box0 }

\def\xara(#1,#2,#3,#4){\left(\matrix{#1 & #2\cr #3 & #4\cr}\right)}

\def\w{\omega}

\def\six_j(#1,#2,#3,#4,#5,#6){\left\{\matrix{#1 & #2 & #3\cr
                                         #4 & #5 & #6\cr}\right\}}
\def\nine_j(#1,#2,#3,#4,#5,#6,#7,#8,#9){\left\{\matrix{#1 & #2 & #3\cr
                                        #4 & #5 & #6\cr
                                         #7 & #8 & #9\cr}\right\}}

\def\Ener(#1,#2){ \sqrt{{#1}^2+{#2}^2} }

\def\overlay#1#2{\setbox0=\hbox{$#1$}\setbox1=\hbox to \wd0{\hss$#2$\hss}#1%
\hskip -1\wd0\copy1}

\def\bold#1{\setbox0=\hbox{$#1$}%
      \kern-.025em\copy0\kern-\wd0
      \kern.05em\copy0\kern-\wd0
      \kern-.025em\raise.0433em\box0 }

\def\S11{S_{11}(1535)}
\def\E0+{E_{0^+}}

\def\footnoterule{\kern-3pt \hrule width \hsize \kern2.6pt}

\newcommand{\beq}{\begin{equation}}
\newcommand{\eeq}{\end{equation}}
\newcommand{\ba}{\begin{array}}
\newcommand{\ea}{\end{array}}
\newcommand{\beqa}{\begin{eqnarray}}
\newcommand{\eeqa}{\end{eqnarray}}
\newcommand{\bd}[1]{ \mbox{\boldmath $#1$}  }

\newcommand{\Pbar}{\not{\!P}}

\newcommand{\phibar}{\not{\!{\varphi}}}

\newcommand{\be}{\begin{equation}}
\newcommand{\ee}{\end{equation}}

\newcommand{\nsigma}{\mbox{\boldmath $\sigma$}}

\newcommand{\np}{\mbox{\boldmath $p$}}
\newcommand{\nq}{\mbox{\boldmath $q$}}
\newcommand{\nr}{\mbox{\boldmath $r$}}
\newcommand{\ns}{\mbox{\boldmath $s$}}
\newcommand{\nn}{\mbox{\boldmath $n$}}
\newcommand{\nl}{\mbox{\boldmath $l$}}

\begin{document}
\begin{titlepage}
\mbox{}
\vspace*{2.5\fill}
\begin{center}
{\Large \bf A$(\vec{e},e'\vec{p}\, )$B responses: from bare nucleons to complex nuclei}\\
\vspace{1\fill}
\end{center}
\begin{center}
{\large Javier R. Vignote$^{1}$, M.C. Mart\'{\i}nez$^{2}$,
J.A. Caballero$^{2}$, E. Moya de Guerra$^{3}$, and J.M. Udias$^{1}$}
\end{center}
\begin{small}
\begin{center}
$^{1}${\sl Departamento de F\'{\i}sica At\'omica, Molecular y Nuclear \\
        Universidad Complutense de Madrid,
        E-28040 Madrid, Spain}\\[2mm]
$^{2}${\sl Departamento de F\'{\i}sica At\'omica, Molecular y Nuclear \\ 
Universidad de Sevilla, Apdo. 1065, E-41080 Sevilla, Spain }\\[2mm]
$^{3}${\sl Instituto de Estructura de la Materia, 
        CSIC, 
         Serrano 123, E-28006 Madrid, Spain }
\end{center}
\end{small}
\kern 1. cm \hrule \kern 3mm
\begin{small}
\noindent
{\bf Abstract}
\vspace{3mm}

We study the occurrence of factorization in polarized and unpolarized 
observables in coincidence quasi-elastic electron scattering. Starting with the
relativistic distorted wave impulse approximation, we reformulate the effective
momentum approximation and show that the latter leads to observables which 
factorize under some specific conditions. Within this framework, the role 
played by final state interactions and, in particular, by the spin-orbit term 
is explored. Connection with the nonrelativistic formalism is studied in 
depth. Numerical results are presented to illustrate the analytical 
derivations and to quantify the differences between factorized and 
unfactorized approaches.

\kern 2mm

\end{small}
 
\kern 2mm \hrule \kern 1cm
\end{titlepage}

\section{Introduction}
 
Quasielastic $(e,e'p)$ reactions have provided over the years an enormous 
wealth of information on nuclear structure, particularly, on single particle
degrees of freedom: energies, momentum distributions and spectroscopic factors 
of nucleons inside nuclei~\cite{Bof96,Kel96,Frul85}. In recent years important 
efforts have been devoted to provide more realistic theoretical descriptions 
of these processes~\cite{Pick87,Pick89,Udi93,Udi95,Udi96,Udi99,Udi00,Udi01,Far03,Meucci01,Meucci02,Ryck99,Cris04}. However, there are still uncertainties 
associated to the various ingredients that enter in the reaction dynamics: 
final state interactions (FSI), off-shell effects, nuclear correlations,
relativistic degrees of freedom or meson exchange currents (MEC). These
ingredients affect the evaluation of electron scattering observables and
hence lead to ambiguities in the information on the nuclear and nucleon 
structure that can be extracted from experiments.
In recent years, electron beam polarization as well as  polarization
degrees of freedom for the outgoing nucleon can be measured, what makes it 
possible to
extract a new wealth of observables from quasielastic $(\vec{e},e'\vec{p})$
reactions. For instance ratios of transferred polarizations are used
to measure ratios of nucleon form factors.

One of the basic results which has made
$(e,e'p)$ reactions so appealing for investigations of single particle
properties is the factorized approach~\cite{Bof96,Cab93,Cab98}.
Within this approximation, the $(e,e'p)$ differential cross section factorizes
into a single-nucleon cross section, describing electron proton scattering,
and a spectral function which gives the probability to find a proton in the
target nucleus with selected values of energy and momentum compatible with the
kinematics of the process. The simplicity of the factorized result makes
it possible to get a clear image of the physics contained in the problem. 
Even being known that factorization does not hold in general,
it is often assumed that the breakdown of factorization is not too severe, and
then it is still commonplace to use factorized calculations for few
body systems or for inclusive scattering.
The importance of factorization lies on the fact that the interpretation of
experimental data  is still usually based on this property
by defining an {\sl effective} spectral  function that is extracted from 
experiment in the form of a reduced cross section. Assuming that factorization 
holds at least approximately, reduced cross section would yield information on
momentum distributions of the nucleons inside the nucleus.
On the other hand, these momentum distributions would cancel
when taking  ratios of cross sections and consequently these ratios
might give information on the electromagnetic form factors of the 
nucleons~\cite{Die01,Str03}. 
 
In spite of the importance of the factorization assumption, there have been 
however almost no formal (and very few  quantitative) studies of its validity. 
So far, it has been shown by different authors \cite{Frul85,Cab98,Cris1} that
in the nonrelativistic case  and when using plane waves to describe the ejected 
nucleon (PWIA), factorization holds exactly for the 
{\em unpolarized cross section}. When interactions in the final state are 
included (DWIA), then certain further assumptions are  needed to recover the 
factorized result \cite{Frul85}. The meaning and importance of the additional 
assumptions required to attain a factorized result has not been quantitatively 
studied thoroughly.
 
In the relativistic case,  factorization of the unpolarized
cross section is broken even without FSI, due to  the negative energy 
components  of the bound nucleon wave function ~\cite{Cab98,Cris1}.
A quantitative estimate of the breakdown of factorization is lacking for the 
relativistic case when taking into account FSI.

Furthermore, there has not been any study of the validity of the
factorization picture for polarization observables, even though
this factorized picture is implicitly assumed when using
ratios of transferred polarizations to determine nucleon
form factors \cite{Die01,Str03}.
 
Within a nonrelativistic framework, the breakdown of factorization has been
usually interpreted as due to the spin-orbit dependent optical potentials.
We note however, that other effects such as the Coulomb distortion of the
electron waves, and contributions beyond the impulse approximation (IA) such as
MEC, play also a role in breaking factorization.
In the particular case of the plane wave limit (i.e.,
neglecting FSI between the ejected proton and the residual nucleus)
factorization is strictly satisfied in IA at the level of the transition
amplitude~\cite{Frul85,Cab98}. This contrasts strongly with the
relativistic formalism, where the enhancement of the lower components of the
bound nucleon wave function destroys factorization of the transition amplitude 
even in the
case of no FSI. Hence, an important difference between relativistic and
nonrelativistic approaches already emerges in the plane wave limit. Whereas
factorization holds in nonrelativistic PWIA, it does not in the relativistic 
plane wave  impulse approximation (RPWIA), which includes negative-energy 
components in the bound nucleon wave function~\cite{Cab98,Cris1}.

As mentioned above, the mechanism that breaks factorization has been only 
established for the unpolarized cross section in the nonrelativistic 
approach. Here we explore such mechanisms for both polarized and unpolarized 
observables starting from the more complex relativistic distorted wave 
impulse approximation (RDWIA) and making simplifying assumptions that lead to 
factorization. We make also
the connection with the nonrelativistic framework and present conclusions
that are valid in both relativistic and nonrelativistic cases.
It is important to point out that most of the $(e,e'p)$ experiments 
performed recently involved energies and momenta high enough to make 
compulsory the use of relativistic nucleon dynamics. Within this context, the 
RDWIA, which incorporates kinematical and dynamical relativistic effects, 
has proved its 
capability to explain polarized and unpolarized $(e,e'p)$ experimental 
data~\cite{Udi93,Udi99,Udi00,Udi01}. Starting from the RDWIA, 
the effective momentum approximation (EMA-noSV), originally introduced by 
Kelly~\cite{Kel97}, is reformulated here paying special attention to aspects 
concerned with the property of factorization. In addition, an analysis is made 
of the 
various assumptions that lead to {\sl factorized} polarized and unpolarized 
observables and which are mainly linked to the spin-orbit dependence of the 
problem. Finally, a quantitative estimate of the validity (or  breakdown) of 
factorization is made for different observables that are commonly  extracted 
from $(e,e'p)$ experiments.
 
The paper is organized as follows: in Sec.~II we outline the basic RDWIA
formalism and revisit the EMA-noSV approach, emphasizing its connection with
the factorized approximation. In Sec.~III we present our analysis for
polarized and unpolarized observables, deriving the specific conditions which
lead to {\sl factorization}. In Sec.~IV we concentrate on reduced cross 
sections and connect them to the momentum distributions. Results for polarized 
and unpolarized observables are presented in Sec.~V. Numerical calculations 
performed within different approaches are compared. Finally, in Sec.~VI we 
draw our conclusions.

\section{Relativistic distorted wave impulse approximation (RDWIA)}

The RDWIA has been described in detail in previous works 
(see for instance~\cite{Udi93,Udi01}). In this section we limit our attention 
to those aspects needed for later discussion of the results presented. In 
RDWIA the one body nucleon current
\beq
J^\mu(\w,{\bd q})=\int d{\bd p} \, \overline{\psi}_F^{s_F}({\bd p}+{\bd q})
                \widehat{J}^\mu\psi_{\kappa_b}^{\mu_b}({\bd p})\, ,
\label{amplit}
\eeq
where $\omega$ and ${\bd q}$ are the energy and momentum of the exchanged 
virtual photon, is calculated with relativistic $\psi_{\kappa_b}^{\mu_b}$ 
and $\psi_F^{s_F}$ wave functions for initial bound and final outgoing 
nucleons, respectively, and with relativistic nucleon current operator 
$\widehat{J}^\mu$. 

The bound state wave function is a four-spinor with well defined angular 
momentum quantum numbers $\kappa_b$, $\mu_b$ corresponding to the shell under 
consideration. In momentum space it is given by 
\beq
\psi_{\kappa_b}^{\mu_b}({\bd p})=
        \frac{1}{(2\pi)^{3/2}}\int d{\bd r}e^{-i{\bd p}\cdot{\bd r}}
                \psi_{\kappa_b}^{\mu_b}({\bd r})
                        =(-i)^{\ell_b}
        \left(\begin{array}{@{\hspace{0pt}}c@{\hspace{0pt}}}
            g_{\kappa_b}(p) \\
               S_{\kappa_b} f_{\kappa_b}(p)\frac{{\bd \sigma}\cdot{\bd p}}{p}
               \end{array}\right)\Phi_{\kappa_b}^{\mu_b}(\widehat{\np}) \, , 
\label{psiB}
\eeq
which is the eigenstate of total angular momentum $j_b=|\kappa_b|-1/2$,
and $\Phi_{\kappa_b}^{\mu_b}(\widehat{\np})$ are the spinor harmonics
\beq
\Phi_{\kappa_b}^{\mu_b}(\widehat{\np})=\sum_{m_{\ell_b} h}
\langle \ell_b m_{\ell_b} \frac{1}{2} h | j_b \mu_b \rangle
Y_{\ell_b}^{\mu_b}(\widehat{\np})\chi_{\frac{1}{2}}^h \, , 
\eeq
with $\ell_b=\kappa_b$ if $\kappa_b>0$ and $\ell_b=-\kappa_b-1$ if 
$\kappa_b<0$.  

The wave function for the outgoing proton is a solution of the Dirac 
equation containing scalar (S) and vector (V) optical 
potentials~\cite{Udi93,Udi95}. For a nucleon scattered with asymptotic 
momentum $\np_F$ and spin projection $s_F$, its expression is
\beq
\psi_F^{s_F}({\bd p})=4\pi\sqrt{\frac{E_F+M}{2E_F}}\sum_{\kappa \mu m}
e^{-i\delta^\ast_\kappa}i^\ell \langle \ell m \frac{1}{2} s_F|j \mu \rangle
        Y_\ell^{m\ast}(\widehat{\np}_F)\psi_\kappa^{\mu}({\bd p}) \, .
\label{111}
\eeq
As the optical potential may be in general complex the phase shifts and
radial functions are also complex, and the wave function 
$\psi_{\kappa}^{\mu}({\bd p})$ is given by
\beq
\psi_{\kappa}^{\mu}({\bd p})=(-i)^\ell
\left(\begin{array}{@{\hspace{0pt}}c@{\hspace{0pt}}}
                g^\ast_\kappa(p) \\
                 S_\kappa f^\ast_\kappa (p)\frac{{\bd \sigma}\cdot{\bd p}}{p}
                \end{array}\right)\Phi_\kappa^\mu(\widehat{\np}) \, .
\label{112} 
\eeq

Assuming plane waves for the electron (treated in the extreme relativistic 
limit), the differential cross section for outgoing nucleon polarized 
$A(\vec{e},e'\vec{p}\,)B$ reactions can be written in the laboratory system in 
the general form
\beq
\frac{d \, \sigma}{d\varepsilon_f d \, \Omega_f d \, \Omega_F}=
\frac{E_F p_F}{(2\pi)^3} \sigma_{M}f_{rec} \, \w_{\mu \nu}W^{\mu \nu} \, ,
\label{xsection}
\eeq
where $\sigma_M$ is the Mott cross section, $\{\varepsilon_f,\Omega_f\}$ are 
the energy and solid angle corresponding to the scattered electron and 
$\Omega_F = (\theta_F,\phi_F)$ the solid angle for the outgoing proton. 
The factor $f_{rec}$ is the usual recoil factor
$f_{rec}^{-1}=|1-(E_F/E_B)(\np_B \cdot \np_F)/p_F^2|$, being $\np_B$ and 
$E_B$ the momentum and energy of the residual nucleus, respectively.
Finally, $\omega_{\mu\nu}$ is the familiar leptonic tensor that can be 
decomposed into its symmetric (helicity independent) and antisymmetric 
(helicity dependent) parts and $W^{\mu \nu}$ is the hadronic tensor
which contains all of the hadronic dynamics of the process. The latter is 
defined from bilinear combinations of the one body nucleon current matrix 
elements given in Eq.~(\ref{amplit}), as
\beq
W^{\mu\nu}= \frac{1}{2j_b+1}\sum_{\mu_b}
J^{\mu \ast}(\omega,\nq) J^\nu (\omega,\nq) \, .
\eeq

The cross section can be also written in terms of hadronic responses by making 
use of the general properties of the leptonic tensor. For 
$(\vec{e},e'\vec{p}\,)$ reactions with the incoming electron polarized and the 
final nucleon polarization also measured, a total set of eighteen response 
functions contribute to the cross section. Its general expression is written 
in the form 
\begin{eqnarray}
\frac{d\sigma}{d\varepsilon_f d\Omega_f d\Omega_F}
 &=& \frac{E_F p_F}{(2\pi)^3} \sigma_{M}f_{rec}\, \frac{1}{2}
\left\{ v_L\left(R^{L}+R^{L}_{n}\widehat{S}_n\right )+
     v_T\left (R^{T}+R^{T}_n\widehat{S}_n \right) \right. \nonumber \\
&+& \left.
 v_{TL}\left[\left(R^{TL}+R^{TL}_{n}\widehat{S}_n\right)\cos\phi_F +
             \left(R^{TL}_{l}\widehat{S}_l+R^{TL}_{s}\widehat{S}_s\right)\sin\phi_F \right] \right.
\nonumber \\
&+& \left. 
v_{TT}\left[\left( R^{TT}+R^{TT}_{n}\widehat{S}_n\right)\cos2\phi_F +
            \left( R^{TT}_{l}\widehat{S}_l+R^{TT}_{s}\widehat{S}_s\right )\sin 2\phi_F \right] \right.  
\nonumber \\
&+& \left. 
h\left\{v_{TL'}\left[\left(R^{TL'}_{l}\widehat{S}_l+R^{TL'}_{s}\widehat{S}_s\right)\cos\phi_F
+\left(R^{TL'}+R^{TL'}_{n}\widehat{S}_n\right)\sin\phi_F\right] \right. \right.
\nonumber \\
& + & \left. \left.
v_{T'}\left[R^{T'}_{l}\widehat{S}_l+R^{T'}_{s}\widehat{S}_s\right]\right\} \right\} \, ,
\label{responses}
\end{eqnarray}
where $v_\alpha$ are the usual electron kinematical factors~\cite{Pick89,Udi01} 
and $h=\pm 1$ is the incident electron helicity.
The polarized and unpolarized 
nuclear response functions are constructed directly by taking the appropriate 
components of the hadronic tensor $W^{\mu\nu}$ (see Ref.~\cite{Pick89} 
for their explicit expressions). The cross section dependence on the recoil 
nucleon polarization is specified by the components $\widehat{S}_k$ ($k=l,n,s$) of the ejected proton rest frame spin $(\ns_F)_R$ 
along the directions: $\nl = \np_F /p_F$, $\nn = (\nq \times \np_F)/|\nq \times \np_F|$ and $\ns = \nn \times \nl$. 

To finish this section and in order to ease the analysis of the results, 
the cross section can be also expressed in terms of the usual polarization asymmetries, which are given as ratios between different classes of response functions, 
\begin{equation}
\frac{d\sigma}{d\varepsilon_f d\Omega_f d\Omega_F}=
\frac{\sigma_0}{2}\left[1+P_n \widehat{S}_n+P_l \widehat{S}_l
+P_s \widehat{S}_s+ h \left(A+P'_n \widehat{S}_n+P'_l \widehat{S}_l
+P'_s \widehat{S}_s \right) \right] \, ,
\label{difcross1}
\end{equation}
with $\sigma_0$ the unpolarized cross section, $A$ the electron analyzing power, and
$P_k$ ($P'_k$) the induced (transferred) polarizations. 

\subsection{Factorization and effective momentum approximation}

In nonrelativistic PWIA, the $(e,e'p)$ unpolarized cross section factorizes in 
the form
\begin{equation}
\left(\frac{d\sigma}{d\varepsilon_f d\Omega_f d\Omega_F}\right)^{PWIA} = 
E_F p_F f_{rec}\, 
\sigma_{ep} \, N_{NR}(\np_m)
\, ,
\label{facpwia}
\end{equation}
where $\sigma_{ep}$ is the bare electron-proton cross section usually taken as 
$\sigma_{cc1}$ (or $\sigma_{cc2}$) of de Forest~\cite{For83}, 
and $N_{NR}(\np_m)$ is the {\em non relativistic momentum distribution} that 
represents the probability of finding a proton in the target nucleus with 
missing momentum $\np_m$, compatible with the kinematics of the reaction. It 
is well known that the factorized result in Eq.~(\ref{facpwia}) comes from an 
oversimplified description of the reaction mechanism. FSI, as well as Coulomb 
distortion of  the electron wave functions, destroys in general factorization. 
In fact, most current descriptions of exclusive $(e,e'p)$ reactions involve 
unfactorized calculations. However, the simplicity of the factorized result 
makes it very useful to analyze and interpret electron scattering observables 
in terms of single particle properties of bound nucleons. Therefore it is 
common to quote {\em experimental reduced cross section} or {\em effective 
momentum distribution} on the basis of the experimental unpolarized cross 
section as
\begin{equation}
\rho^{exp}(\np_m) = \frac{
\left(\frac{d\sigma}{d\varepsilon_f d\Omega_f d\Omega_F}\right)^{exp}}
{E_F p_F f_{rec}\, \sigma_{ep}} \, . 
\label{facexp}
\end{equation}
A similar expression can be used for the theoretical reduced cross section, 
\begin{equation}
\rho^{th}(\np_m) = \frac{
\left(\frac{d\sigma}{d\varepsilon_f d\Omega_f d\Omega_F}\right)^{th}}
{E_F p_F f_{rec}\, \sigma_{ep}} \, ,
\label{facth}
\end{equation}
constructed from the the theoretical unpolarized $(e,e'p)$ cross section, 
independently of whether it is calculated 
within a relativistic or nonrelativistic formalism. 
We will say that the factorization 
property is satisfied by $\rho^{th}(\np_m)$ when the theoretical unpolarized 
cross section factors out exactly $\sigma_{ep}$, and then, the theoretical 
reduced cross section does not depend on it. 

As we will demonstrate later in this paper, factorization is not a property 
exclusive of the nonrelativistic PWIA approach. It is well known 
that, due to the negative energy components of the bound proton wave function, 
factorization is not satisfied even in RPWIA~\cite{Cab98}. However, if we 
neglect the contribution from the negative energy components, the unpolarized 
cross section factorizes to a similar expression as in Eq.~(\ref{facpwia}).
  
Starting from a fully relativistic calculation of the nuclear current,
in what follows we explore the most general conditions 
under which factorization is 
recovered. First, it is important to note that in order to extract the 
elementary cross section ``$\sigma_{ep}$'' from the general relativistic 
theory (RDWIA), the upper and lower components of the 
relativistic wave functions that enter in Eq.~(\ref{amplit}) must be forced 
to satisfy the ``free'' relationship with momenta determined by asymptotic 
kinematics at the nucleon vertex, that is
\beq
\psi_{down}(\np)=\frac{{\bd \sigma} \cdot \np_{as}}{E_{as}+M}\psi_
{up}(\np) \, ,
\label{asymp}
\eeq
with $E_{as}=\sqrt{\np_{as}^2+M^2}$ and $\np_{as}$ the asymptotic momentum
corresponding to each nucleon. In what follows we discuss this 
condition~(\ref{asymp}) in the nonrelativistic language.

The nonrelativistic formalism is based on bispinors $\chi(\np)$ solutions of 
Schr\"odinger-like equations. Generally, the nonrelativistic formalism can be 
analyzed using the following semirelativistic (SR) four-spinor
\beq 
\psi^{SR}(\np) = \frac{1}{\sqrt{N}}
\left(\begin{array}{@{\hspace{0pt}}c@{\hspace{0pt}}}
                \chi(\np) \\
                \frac{\nsigma \cdot \np}{E+M}\chi(\np)
                \end{array}\right) \, ,
\label{SR}
\eeq 
to be introduced in Eq.~(\ref{amplit}) in order to calculate a 
relativistic-like nucleon current amplitude. In this way the relativistic 
kinematics is fully taken into account and no expansions in $p/M$ are needed.
The one body nucleon current matrix element takes then the following form:
\beq
J^\mu(\w,{\bd q})=\int d{\bd p} \, \chi_F^{s_F \dagger}({\bd p}+{\bd q})
                \widehat{J}\,^\mu_{eff}(\np,\nq) \chi_{j_b}^{\mu_b}({\bd p}) 
\, ,
\label{amplitnonrel}
\eeq
with $\widehat{J}\,^\mu_{eff}(\np,\nq)$ now an effective (2x2) current 
operator that occurs between bispinor wave functions $\chi_F^{s_F}$ 
($\chi_{j_b}^{\mu_b}$) 
for the outgoing (bound) nucleon respectively.   

The calculation of the nuclear amplitude using four-spinors like the one 
written in Eq.~(\ref{SR}), implies removal of the enhancement of the 
lower components that is present in the four-spinors of Eqs.~(\ref{psiB}) and 
(\ref{111}). This is a well known fact present in nonrelativistic 
calculations, but this alone is not enough to get factorization. It is 
also required the use of exactly the same nuclear current operator as in a free 
electron-proton  scattering. In Eq.~(\ref{amplitnonrel}) then, 
the non-truncated effective current operator must be evaluated at the 
asymptotic momentum values, leading to
\beq
J^\mu(\w,{\bd q})=\int d{\bd p} \, \chi_F^{s_F \dagger}({\bd p}+{\bd q})
                \widehat{J}\,^\mu_{eff}(\np_F-\nq,\nq) \chi_{j_b}^{\mu_b}
({\bd p})\, .
\label{amplitnonrelas}
\eeq
One can show that this condition is implicit in 
one of the necessary assumptions introduced in Ref.~\cite{Frul85} to 
recover factorization in the nonrelativistic case.
 
In a relativistic calculation, the assumptions written in 
Eq.~(\ref{asymp}) set up the so-called Effective Momentum Approximation with no 
Scalar and Vector terms 
(EMA-noSV)\footnote{The factorization property could be also analyzed
within the framework of the asymptotic projection approach 
(see Refs.~\cite{Udi99,Udi01,Cris04} for details)}, originally introduced by 
Kelly~\cite{Kel97}, to which we will refer in what follows as EMA. The EMA 
approximation in the relativistic framework, or the 
nonrelativistic calculation based on Eq.~(\ref{amplitnonrelas}), are essentially 
the same conditions which are necessary to recover factorization, in   
either formalism. These conditions are necessary but not sufficient and in what 
follows, we concentrate on the EMA case to study 
additional assumptions needed to obtain factorization.
 
In EMA, the bound nucleon wave function in momentum space is given by
\beqa
& &\psi^{\mu_b \, EMA}_{\kappa_b}(\np)
=(-i)^{\ell_b}
        \left(\begin{array}{@{\hspace{0pt}}c@{\hspace{0pt}}}
                 g_{\kappa_b}(p) \\
                 \frac{{\bd \sigma}\cdot{\bd p}_I}{E_I+M}g_{\kappa_b}(p)
                \end{array}\right)
              \Phi_{\kappa_b}^{\mu_b}(\widehat{\np}) \, ,
\eeqa
with $E_I=\sqrt{{\bd p}_I^2+M^2}$ and ${\bd p}_I={\bd p}_F-{\bd q}$.
Likewise the outgoing relativistic distorted wave function in Eq.~(\ref{112}) 
becomes
\beq
\psi_{\kappa}^{\mu \, EMA}({\bd p})=(-i)^\ell
\left(\begin{array}{@{\hspace{0pt}}c@{\hspace{0pt}}}
                g^\ast_\kappa(p) \\
                 \frac{{\bd \sigma}\cdot{\bd p}_F}{E_F+M}g^\ast_\kappa(p)
                \end{array}\right)\Phi_\kappa^\mu(\widehat{\np}) \, .
\label{112ema}
\eeq

Introducing these expressions into the equation of the one body nucleon 
current matrix element~(Eq.~(\ref{amplit})), we get
\beqa
J^\mu_{EMA} &=&
\sum_{s h}
\left[\overline{u}(\np_F,s)\widehat{J}^\mu u(\np_I,h)\right]
\sum_{\kappa \mu m}
\langle \ell m \frac{1}{2} s_F | j \mu \rangle
Y_\ell^{m \ast}(\widehat{\np}_F)
\nonumber \\
&\times &
\sum_{m_{\ell_b} m_{\ell}}
\langle \ell_b m_{\ell_b} \frac{1}{2} h | j_b \mu_b \rangle
\langle \ell m_{\ell} \frac{1}{2} s | j \mu \rangle
U_{\kappa_b \, m_{\ell_b}}^{\kappa \, m_{\ell}}(\np_F,\nq)
\nonumber \\
&\equiv &
\sum_{s h} J^{\mu}_{bare}(\np_F s,\np_I h) A^{\mu_b}_{s h}(\np_F,\nq) \, ,
\label{Jema}
\eeqa
where we have written both nucleon wave functions in terms of free positive 
energy Dirac spinors and we have introduced the bare nucleon current matrix 
element
\beq
J^{\mu}_{bare}(\np_F s,\np_I h)=
\overline{u}(\np_F,s)\widehat{J}^\mu u(\np_I,h) \,, 
\eeq
with the term $U_{\kappa_b \, m_{\ell_b}}^{\kappa \, m_{\ell}}(\np_F,\nq)$
given by
\beqa
U_{\kappa_b \, m_{\ell_b}}^{\kappa \, m_{\ell}}=
\frac{8\pi M}{\sqrt{2E_F(E_I+M)}}(-i)^{\ell_b}
\int d \np \, g_{\kappa_b}(p)
g_{\kappa}^\ast(|\np +\nq |)
Y_{\ell_b}^{m_{\ell_b}}(\widehat{\np})Y_\ell^{m_{\ell} \ast}(\widehat{\np+\nq})
e^{i\delta_{\kappa}} \, ,
\eeqa
and the amplitude
\beqa
A^{\mu_b}_{sh}(\np_F,\nq)=
\sum_{\kappa \mu m}
\langle \ell m \frac{1}{2} s_F | j \mu \rangle
Y_\ell^{m \ast}(\widehat{\np}_F)
\sum_{m_{\ell_b} m_{\ell}}
\langle \ell_b m_{\ell_b} \frac{1}{2} h | j_b \mu_b \rangle
\langle \ell m_{\ell} \frac{1}{2} s | j \mu \rangle
U_{\kappa_b \, m_{\ell_b}}^{\kappa \, m_{\ell}}(\np_F,\nq) \, .
\eeqa

The result in Eq.~(\ref{Jema}) defines the nucleon current in EMA, and 
is our starting point for the analysis of the conditions that may lead to 
{\sl factorized} observables. Notice that $J^{\mu}_{EMA}$ involves a sum over 
initial and final spin projections ($s$, $h$) of the bare nucleon current, 
times an amplitude that depends on the bound and ejected nucleon wave 
functions. Factorization in $J^{\mu}_{EMA}$ occurs if 
$A^{\mu_b}_{s h}(\np_F,\nq)$ does not depend on the spin variables $s$ and $h$. 

Before entering into a detailed discussion of the observables, it is important 
to stress again that factorization may only be achieved 
assuming EMA and/or asymptotic projection, i.e., neglecting dynamical 
enhancement of the lower components in the nucleon wave functions. This is  
a priori assumed within some nonrelativistic calculations.


\section{Analysis of observables within EMA}


In this section we investigate  the conditions that lead to factorization of 
polarized and unpolarized observables. Response functions, 
transverse-longitudinal asymmetry, electron analyzing power, as well as 
induced and transferred polarizations are examined. 
The analysis is made directly at the level of the hadronic tensor 
which, within the EMA approach, can be written in the following way:
\beqa
W^{\mu\nu}_{EMA} &=& \frac{1}{2j_b+1}\sum_{\mu_b}
\left(J^\mu_{EMA}\right)^\ast
J^\nu_{EMA}
\nonumber \\
&=&
\sum_{ss'}\sum_{hh'}
\left[J^{\mu}_{bare}(\np_F s,\np_I h)\right]^\ast
J^{\nu}_{bare}(\np_F s',\np_I h')
\nonumber \\
& \times &
\frac{1}{2j_b+1}\sum_{\mu_b}
\left[A^{\mu_b}_{s h}(\np_F,\nq)\right]^\ast
A^{\mu_b}_{s' h'}(\np_F,\nq) \, .
\label{tensorema}
\eeqa
Note that in Eq.~(\ref{tensorema}) $s,s'$ are the spin variables corresponding 
to the outgoing nucleon, while $h,h'$ correspond to the bound nucleon.

In order to simplify the analysis that follows, the general expression of the 
hadronic tensor can be written in a more compact form as
\beq
W^{\mu\nu}_{EMA}=\sum_{ss'}\sum_{hh'}{\cal W}^{\mu\nu}_{ss',hh'}
X_{ss',hh'}^{s_F}(\np_F,\nq) \, , 
\label{wmunu}
\eeq
where we have introduced a general bare-nucleon tensor 
${\cal W}^{\mu\nu}_{ss',hh'}$,
\be
{\cal W}^{\mu\nu}_{ss',hh'}=
\left(J^{\mu}_{bare}\right)^\ast
J^{\nu}_{bare}=
\left[\overline{u}(\np_F,s)\widehat{J}^\mu u(\np_I,h)\right]^\ast
\left[\overline{u}(\np_F,s')\widehat{J}^\nu u(\np_I,h')\right]  \, ,
\label{calwmunu}
\ee     
and a general spin dependent momentum distribution function $X_{ss',hh'}^{s_F}$,
\beqa
X_{ss',hh'}^{s_F}(\np_F,\nq) &=&
\frac{1}{2j_b+1}\sum_{\mu_b}
\left[A^{\mu_b}_{s h}(\np_F,\nq)\right]^\ast
A^{\mu_b}_{s' h'}(\np_F,\nq)
\nonumber \\
&= &
\frac{1}{2j_b+1}\sum_{\mu_b}
\sum_{\kappa \mu m} \sum_{\kappa' \mu' m'}
\langle \ell m \frac{1}{2} s_F | j \mu \rangle
\langle \ell' m' \frac{1}{2} s_F | j' \mu' \rangle
Y_\ell^{m}(\widehat{\np}_F) Y_{\ell'}^{m' \ast}(\widehat{\np}_F)
\nonumber \\
&\times &
\sum_{m_{\ell_b} m_{\ell}}
\sum_{m_{\ell_b}' m_{\ell}'}
\langle \ell_b m_{\ell_b} \frac{1}{2} h | j_b \mu_b \rangle
\langle \ell m_{\ell} \frac{1}{2} s | j \mu \rangle
\langle \ell_b m_{\ell_b}' \frac{1}{2} h' | j_b \mu_b \rangle
\langle \ell' m_{\ell}' \frac{1}{2} s' | j' \mu' \rangle
\nonumber \\
&\times &
U_{\kappa_b \, m_{\ell_b}}^{\kappa \, m_{\ell} \ast}(\np_F,\nq)
\, U_{\kappa_b \, m_{\ell_b}'}^{\kappa' \, m_{\ell}'}(\np_F,\nq) \, .
\label{momdis}
\eeqa

Making use of general symmetry properties (see Appendix A), the bare-nucleon 
tensor in Eq.~(\ref{calwmunu}) can be decomposed into terms which are
symmetric and antisymmetric under interchange of $\mu$ and $\nu$. Each of 
these terms shows a different dependence on the spin variables: 
$ss'$ and/or $hh'$. Explicitly, the bare nucleon tensor can be written in the 
form
\be
{\cal W}^{\mu\nu}_{ss',hh'} = {\cal S}^{\mu\nu} \, \delta_{ss'}
\, \delta_{hh'}+
{\cal A}^{\mu\nu}_{hh'} \, \delta_{ss'}+{\cal A}^{\mu\nu}_{ss'} \,
\delta_{hh'} + {\cal S}^{\mu\nu}_{ss', \, hh'} \, ,
\label{calwmunu2}
\ee
where ${\cal S}$ (${\cal A}$) refers to symmetric (antisymmetric) tensors.
Notice that the first (symmetric) term in Eq.~(\ref{calwmunu2}) does not 
depend on the initial bound neither on the final outgoing nucleon spin 
variables; the antisymmetric second (third) term depends solely on the 
initial (final) spin projections; finally, the fourth (symmetric) term 
presents dependence on both initial and final nucleon spin projections 
simultaneously. This bare-nucleon tensor would lead to the
$\sigma_{ep}$ cross section in Eq.~(\ref{facpwia}). 

The general result for the bare nucleon tensor given in Eq.~(\ref{calwmunu2}) 
constitutes the starting point for the analysis of factorization for polarized 
as well as unpolarized observables. In what follows we explore the specific 
conditions, linked to the spin dependence in the problem, that lead to 
{\sl factorized} results. We investigate separately the role played by the 
dependence on the initial and/or final nucleon spin variables. As we show in 
next subsections, the factorization property at the level of spin-averaged 
squared matrix elements is intimately connected with the spin dependence: 
a bound nucleon in a s-wave or, in general, no spin-orbit coupling effects on 
the radial nucleon wave functions, may lead for some specific observables 
to exactly {\sl factorized} results. As it is clear from the analogy between 
Eq.~(\ref{amplitnonrelas}) and Eq.~(\ref{amplit}) with the input from 
Eq.~(\ref{asymp}), the analysis of 
spin dependence here and in what follows is also valid for the nonrelativistic 
case.

\subsection{No spin-orbit in the initial state}

The general expression of $X_{ss',hh'}^{s_F}$ (\ref{momdis}) is greatly 
simplified for no spin-orbit in the initial state or, more generally in 
LS coupling. For instance in the case of nucleon knockout from s-shells the 
orbital angular momentum $\ell_b=0$ and the spin dependent momentum 
distribution is simply given by
\be 
X_{ss',hh'}^{s_F}(\np_F,\nq) = N_{ss'}^{s_F}(\np_F,\nq) \, \delta_{hh'} \, ,
\label{nolsf} 
\ee
with
\beqa
N_{ss'}^{s_F}(\np_F,\nq)&=&
\frac{1}{2j_b+1}
\sum_{\kappa \mu m} \sum_{\kappa' \mu' m'}
\langle \ell m \frac{1}{2} s_F | j \mu \rangle
\langle \ell' m' \frac{1}{2} s_F | j' \mu' \rangle
Y_\ell^{m}(\widehat{\np}_F) Y_{\ell'}^{m' \ast}(\widehat{\np}_F)
\nonumber \\
&\times &
\sum_{m_{\ell} m_{\ell}'}
\langle \ell m_{\ell} \frac{1}{2} s | j \mu \rangle
\langle \ell' m_{\ell}' \frac{1}{2} s' | j' \mu' \rangle
U_{-1 \, 0}^{\kappa \, m_{\ell} \ast}(\np_F,\nq)
\, U_{-1 \, 0}^{\kappa' \, m_{\ell}'}(\np_F,\nq) \, .
\eeqa
In the case of no spin-orbit coupling with $\ell_b \neq 0$ waves, a similar 
reduction to Eq.~(\ref{nolsf}) follows after summation of the spin dependent 
momentum distribution $X$ on $j_b=\ell_b \pm 1/2$.

Making use of Eqs.~(\ref{calwmunu2}) and (\ref{nolsf}), the hadronic tensor in 
EMA becomes
\beqa
W^{\mu\nu}_{EMA}&=&\sum_{ss'}N_{ss'}^{s_F}\left[\sum_h
{\cal W}^{\mu\nu}_{ss',hh}\right]
\nonumber \\
&=&
\sum_{ss'}N_{ss'}^{s_F}\left[{\cal S}^{\mu\nu} \, \delta_{ss'}+
{\cal A}^{\mu\nu}_{ss'} \right] = {\cal S}^{\mu\nu} \sum_{s}N_{ss}^{s_F}+
\sum_{ss'}N_{ss'}^{s_F}{\cal A}^{\mu\nu}_{ss'} \, .
\label{nolb}
\eeqa
From this result it clearly emerges that those responses coming from 
the symmetric tensor ${\cal S}^{\mu\nu}$ factorize, while the ones 
coming from the antisymmetric part do not. Let us signal out more precisely
what factorization really means in this situation.

First, note that the momentum distribution function $\sum_sN_{ss}^{s_F}$ that 
multiplies the symmetric tensor depends on the outgoing nucleon spin $s_F$. 
In the case when recoil nucleon polarization is not measured, an extra sum in 
$s_F$ has to be carried out and hence the momentum 
distribution, which is independent of $s_F$, gives rise to the unpolarized 
responses $R^L$, $R^T$, $R^{TL}$ and $R^{TT}$ in Eq.~(\ref{responses}). On 
the other hand, if the spin of the outgoing proton is measured via a 
polarimeter placed along a fixed direction ($\nl$, $\nn$ or $\ns$), the 
momentum distribution, now dependent on the final spin, contributes to the 
induced polarized responses: $R^L_n$, $R^T_n$, $R^{TL}_{n,l,s}$ and
$R^{TT}_{n,l,s}$. Hence, in the case of no spin-orbit coupling 
in the initial bound state, both types of responses (unpolarized and 
induced polarized) factorize, but each kind of response factorizes with a 
different momentum distribution function. Then, the induced polarization 
asymmetries $P_k$ ($k=l,n,s$), which are basically given by the ratio between 
the induced polarized responses $R^\alpha_k$ and the unpolarized ones 
$R^\alpha$, will 
differ from the bare result. 
On the contrary, the momentum distribution functions cancel when 
taking a ratio between two responses of the same kind, i.e., a ratio between
two induced polarized responses along a specific direction, or a ratio 
between two unpolarized responses. Therefore such ratios would coincide with 
the bare results. This property 
can be expressed in the general form
\be
\frac{R^\alpha_k}{R^{\beta}_k}=
\frac{R^\alpha}{R^{\beta}}=\frac{{\cal R}^\alpha}{{\cal R}^{\beta}} \, ,
\label{pp}
\ee 
where $\alpha,\beta=L,T,TL$ or $TT$ and $k=l,n,s$ fixes the recoil nucleon 
polarization direction. The functions ${\cal R}^{\alpha,\beta}$ represent the 
bare-nucleon responses, also usually named single-nucleon 
responses~\cite{Cris1}. The result in Eq.~(\ref{pp}) explains also why the 
$A_{TL}$ asymmetry, which is obtained from the difference of electron 
unpolarized cross sections measured at $\phi_F=0^{\circ}$ and 
$\phi_F=180^{\circ}$ divided 
by the sum, is identical to the bare asymmetry in this case. 
In terms of response functions 
we may write
\beqa
A_{TL}=\frac{v_{TL} R^{TL}}{v_L R^L+v_T R^T+v_{TT} R^{TT}}=
\frac{v_{TL} {\cal R}^{TL}}{v_L {\cal R}^L+v_T {\cal R}^T+v_{TT} {\cal R}^{TT}}
=A_{TL}^{bare} \, .
\label{atlbare}
\eeqa   

To complete the discussion, we note that the electron analyzing power and 
transferred polarization asymmetries involve responses coming from the 
antisymmetric part of the tensor (\ref{nolb}), which do not factorize, 
divided by unpolarized responses obtained from the symmetric tensor term. 
Therefore the behaviour of $A$ and $P'_k$ will differ from the bare one. The 
amount of discrepancy between the factorized and unfactorized calculations of 
different observables is discussed in Sec.~V.

\subsection{No spin-orbit in the final state}

Let us consider now the case of no spin-orbit coupling effects on the radial 
wave function of the outgoing proton. In this case neither $\delta_{\kappa}$ 
nor $g_{\kappa}$ in Eqs.~(\ref{111}), (\ref{112ema}) depend on $j$. 
After some algebra (see Appendix B for details), this condition leads to 
$s=s'=s_F$ in the bare-nucleon tensor, and therefore the momentum distribution 
depends only on the $hh'$ spin variables of the initial nucleon. The hadronic 
tensor is then given by
\be
W^{\mu\nu}_{EMA}=\sum_{hh'}{\cal W}^{\mu\nu}_{s_F s_F,hh'}
\widetilde{N}_{hh'}(\np_F,\nq) \, ,
\label{wmununolsf}
\ee 
where the momentum distribution function $\widetilde{N}_{hh'}(\np_F,\nq)$ is 
defined in Eq.~(\ref{ntilde}) of Appendix B.
Using the decomposition in Eq.~(\ref{calwmunu2}), we can write the following 
expression
\be
W^{\mu\nu}_{EMA}=\left[{\cal S}^{\mu\nu}+{\cal A}^{\mu\nu}_{s_Fs_F}\right]
\sum_h \widetilde{N}_{hh}+\sum_{hh'}\left[
{\cal S}^{\mu\nu}_{s_Fs_F,hh'}+{\cal A}^{\mu\nu}_{hh'}\right]
\widetilde{N}_{hh'}.
\label{wmununolsf2}
\ee

The analysis of how polarized or unpolarized responses behave with respect to 
factorization emerges straightforwardly from Eq.~(\ref{wmununolsf2}). Let us 
discuss each case separately:
\begin{itemize}
\item Unpolarized responses: $R^L$, $R^T$, $R^{TL}$ and $R^{TT}$. They do not 
depend on spin and come from the symmetric part of the tensor, i.e., they are 
given by ${\cal S}^{\mu\nu}\sum_h \widetilde{N}_{hh}$, and hence factorize 
exactly. This result coincides with that one obtained in the nonrelativistic 
study of Ref.~\cite{Frul85}.
\item Transferred polarization responses: $R^{T'}_{l,s}$ and $R^{TL'}_{l,s,n}$. 
 They come from the antisymmetric part of the tensor and depend on the final 
proton spin polarization, i.e., 
${\cal A}^{\mu\nu}_{s_Fs_F}\sum_h \widetilde{N}_{hh}$, in exactly the same form
as displayed in Eq.~(\ref{calwmunu2}). Consequently, these responses also 
factorize.
\item Fifth response $R^{TL'}$. It comes from the antisymmetric part of the 
tensor and does not depend on the recoil nucleon spin, i.e., it is given by 
$\sum_{hh'}\widetilde{N}_{hh'}{\cal A}^{\mu\nu}_{hh'}$ and clearly does not 
factorize.
\item Induced polarized responses: $R^L_n$, $R^T_n$, $R^{TL}_{n,l,s}$ and 
$R^{TT}_{n,l,s}$. They come from the symmetric tensor part and depend 
explicitly on the spin polarization of the outgoing proton, i.e., they are 
constructed from $\sum_{hh'}\widetilde{N}_{hh'}{\cal S}^{\mu\nu}_{s_Fs_F,hh'}$,
 and consequently do not factorize.
\end{itemize}

Once the behaviour of the response functions is established, the asymmetries 
and polarization ratios can be easily analyzed. The case of $A_{TL}$, which 
only depends on the unpolarized responses, reduces to $A_{TL}^{bare}$ 
(see Eq.~({\ref{atlbare})). A similar comment applies also to the 
transferred nucleon polarizations $P'_l$, $P'_s$ and $P'_n$. 
Notice that the momentum distribution function involved in the unpolarized and 
transferred polarized  responses is the same and hence, it cancels when 
forming the polarization ratios. The electron analyzing power $A$ and induced 
asymmetries $P_k$, given in terms of responses which do not factorize, should 
differ from the bare calculations.

As a particular case of no spin-orbit in the final nucleon wave function, it 
is worth to explore the plane wave limit for the outgoing nucleon. In this case (see Eq.~(\ref{ntildpw}) in Appendix B), the momentum distribution 
$\widetilde{N}^{PW}_{hh'}$ is diagonal and independent on $h$, thus the fifth 
response $R^{TL'}$ vanishes since $\sum_{h}{\cal A}^{\mu \nu}_{h h}  = 0$. 
Similarly, the induced polarization responses do not contribute because 
$\sum_{h}{\cal S}^{\mu \nu}_{s_Fs_F,hh}=0$.

\subsection{No spin-orbit in both initial and final states}

To finish with this analysis, let us consider the case of no spin-orbit 
coupling in the initial nor in the final state. In this situation, 
factorization already comes out at the level of the nuclear current matrix 
element. Note that $\ell_b=0$ in Eq.~(\ref{jemanolsf}) of Appendix B,
leads to $h=\mu_b$ and the matrix element simply reads
\beq
J_{EMA}^\mu=\overline{u}(\np_F,s_F)\widehat{J}^\mu u(\np_I,\mu_b)
\, U_{-1}^0(\np_F,\nq) \, ,
\label{jumu}
\eeq
where $U_{-1}^0$ is defined in Eq.~(\ref{utilde}). This result resembles the
situation occurring in the free case. From the current (\ref{jumu})
the hadronic tensor can be written in the form
\beq
W^{\mu \nu}_{EMA}=\frac{1}{2}\left|U_{-1}^0(\np_F,\nq)\right|^2
\sum_{\mu_b}{\cal W}^{\mu \nu}_{s_Fs_F, \mu_b \mu_b}=
\frac{1}{2} \left|U_{-1}^0(\np_F,\nq)\right|^2\left(
{\cal S}^{\mu \nu}+{\cal A}^{\mu \nu}_{s_F s_F}\right) \, .
\eeq
Then all responses (polarized and unpolarized) factorize with the same 
momentum distribution. Note also that the whole dependence on the nucleon
polarization $s_F$ is contained in the antisymmetric
tensor. This implies that the polarized induced responses must be zero. 
Furthermore, since $\sum_{s_F}{\cal A}^{\mu \nu}_{s_F s_F}  = 0$ 
the unpolarized fifth response $R^{TL'}$ also vanishes.

\section{Reduced cross sections and momentum distributions}

Starting from a shell model approach, the relativistic (vector) momentum 
distribution is defined as follows:
\beq
 N(p_I)=\frac{1}{2j_b+1} \sum_{\mu_b} 
 \psi^{\mu_b \dagger}_{\kappa_b}(\np_I) \psi^{\mu_b}_{\kappa_b}(\np_I) =
\frac{1}{4\pi} \left[g_{\kappa_b}^2(p_I)+f_{\kappa_b}^2(p_I)\right] \, . 
\label{momrel}
\eeq
Using the EMA approximation means projecting out the negative energies 
components of the bound proton wave function, obtaining then the relativistic 
EMA momentum distribution:
\beq
 N^{EMA}(p_I)=\frac{1}{2j_b+1} \sum_{\mu_b} 
\psi^{\mu_b EMA \dagger}_{\kappa_b}(\np_I) \psi^{\mu_b EMA}_{\kappa_b}
(\np_I) = \frac{1}{4\pi} \, \frac{2E_I}{E_I+M} g_{\kappa_b}^2(p_I)\, ,
\label{momemarel}
\eeq 
this expression reduces to the nonrelativistic momentum 
distribution in the proper limit because of its lack of contribution from 
negative energies. 

In general, in a nonrelativistic formalism, the momentum distribution is 
defined from bispinors $\chi_{j_b}^{\mu_b}(\nr)$, solutions of  
Schr\"odinger-like equation:
\beq
 N_{NR}(p_I)=\frac{1}{2j_b+1} \sum_{\mu_b}
\chi^{\mu_b \dagger}_{j_b}(\np_I) \chi^{\mu_b}_{j_b}(\np_I) \, ,
\label{momnonrel}
\eeq 
with $\chi^{\mu_b}_{j_b}(\np_I)$ the Fourier transform of 
$\chi_{j_b}^{\mu_b}(\nr)$,
\beq
\chi^{\mu_b}_{j_b}(\np_I) = \frac{1}{(2\pi)^{3/2}} \int d\nr 
\, e^{-i \np_I \cdot \nr} \, \chi_{j_b}^{\mu_b}(\nr) \, .
\label{fourier}
\eeq

Now, in nonrelativistic PWIA, the wave function for the ejected proton in the
$\nr$-space is
\beq
\chi_F^{s_F PW}(\nr)= 
e^{i \np_F \cdot \nr} \, \chi_{\frac{1}{2}}^{s_F} \, ,
\eeq
and looking at the the Fourier transform in 
 Eq.~(\ref{fourier}), it is natural to define a nonrelativistic distorted 
wave amplitude as follows:
\beq
\chi_{DW}(\np_F, \nq) \equiv \frac{1}{(2\pi)^{3/2}} \int d\nr \,
\chi_F^{s_F \dagger}(\nr) \, e^{i \nq \cdot \nr} \, \chi_{j_b}^{\mu_b}(\nr) \, .
\label{nrdwamp}
\eeq
Two observations are worth mentioning: 
\begin{enumerate}
\item $\chi_{DW}(\np_F, \nq)$ is an amplitude, not a bispinor. 
\item If the final proton wave function is a plane wave, the following 
relationship is satisfied:
\beq
\sum_{s_F}\left|\chi_{PW}(\np_F, \nq) \right|^2 = 
\chi^{\mu_b \dagger}_{j_b}(\np_I) \chi^{\mu_b}_{j_b}(\np_I) \, .
\eeq
\end{enumerate}
So, we can define a nonrelativistic distorted momentum distribution
\beq
\rho_{DW}^{NR}(\np_F,\nq)=\frac{1}{2j_b+1} \sum_{\mu_b} 
\sum_{s_F} 
\left|\chi_{DW}(\np_F, \nq) \right|^2 \, ,
\label{nrdw}
\eeq
that takes into account FSI, and has the property that we recover the 
nonrelativistic momentum distribution in Eq.~(\ref{momnonrel}) in the plane 
wave limit.

Let us generalize the above expression to the relativistic case. We request 
that we recover from it the relativistic EMA momentum distribution 
of Eq.~(\ref{momemarel}) when there is not FSI and the initial 
wave function is evaluated within EMA. For that purpose we define a 
relativistic distorted wave amplitude,
\beq
\psi_{DW}(\np_F, \nq) \equiv \frac{K}{(2\pi)^{3/2}} \int d\nr \,
\psi_F^{s_F \dagger}(\nr) \, e^{i \nq \cdot \nr} \, 
\psi_{\kappa_b}^{\mu_b}(\nr) = \frac{K}{(2\pi)^{3/2}} \int d\np \, 
\psi_F^{s_F \dagger} (\np+\nq) \psi_{\kappa_b}^{\mu_b}(\np)
\label{reldwamp}
\eeq
with $K=\sqrt{(2 E_I E_F)/(E_I E_F + \np_I \cdot \np_F + M^2)}$, so that
the relativistic distorted momentum distribution is given by this amplitude
squared after sum and average over initial and final spins,
\beq
\rho_{DW}(\np_F,\nq)=\frac{1}{2j_b+1} \sum_{\mu_b}
\sum_{s_F}
\left|\psi_{DW}(\np_F, \nq) \right|^2 \, .
\label{reldw}
\eeq
It is easy to check that $\rho_{DW}(\np_F,\nq)$ coincides with the 
relativistic EMA momentum distribution Eq.~(\ref{momemarel}), when one takes 
EMA approximation for the initial wave function and the plane wave limit for 
the final one,  
\beq
\rho_{PW}^{EMA}(\np_F,\nq) = N^{EMA}(p_I) \, . 
\eeq 
It is also important to remark that $\rho_{DW}(\np_F,\nq)$ coincides with the 
corresponding reduced cross section of Eq.~(\ref{facth}) whenever there is 
factorization.

\section{Numerical results}

To show quantitatively the effects introduced by the different 
approaches to the 
general description of $(\vec{e},e'\vec{p})$ reactions, we compare our fully 
RDWIA calculations with the EMA results, exploring also the effects 
introduced by the spin variables in the initial and final nucleon states. 
The results presented in this section illustrate and reinforce the 
conclusions reached in the preceding sections concerning the factorization 
properties.

Guided by the factorization properties one may focus on two different 
aspects in the analysis of observables.
\begin{enumerate}
\item On the one hand, one may factor out the elementary electron-proton 
electromagnetic cross section, in order to isolate and investigate nuclear 
properties like momentum distributions. 

To the extent that factorization holds the 
reduced cross section will follow the momentum distribution. 
In the first part of this section we compare factorized and 
unfactorized results for the reduced cross section to the momentum 
distribution. We show how the different ingredients that break factorization 
may obscure the extraction of momentum distributions. 

First of all, we note that since FSI modify the response of the ejected 
nucleon, it is more adequate to compare reduced cross sections with
distorted momentum distributions (as defined in the previous section). This is 
done in Fig.~\ref{fig1} that we discuss below.

\item On the other hand, one may take ratios between observables to cancel 
out the dependence on the momentum distribution, in order to isolate and 
investigate intrinsic nucleon properties in the nuclear medium, like nucleon 
form factors.  
\end{enumerate}

In Fig.~\ref{fig1} we present reduced cross sections at 
quasielastic kinematics for 
three cases: complete RDWIA approach (solid line), EMA (dashed line) and 
EMA with no spin dependence in the final state, referred to as EMA-noLS 
(dotted line). We also show by a thin solid line the distorted momentum 
distributions ($\rho_{DW}$, Eq.~(\ref{reldw})) which are equivalent to what 
one would obtain from a factorized approach to RDWIA. 
Note that up to $|p_m|$ of around 250 MeV/c, the factorized approach 
$\rho_{DW}$ follows reasonably well the 
full calculation. Actually in this $p_m$ range, EMA and 
EMA-noLS are also reasonable approximations to the complete calculation. 
However, at $|p_m| > 250$ MeV/c the full approach produces more reduced 
cross section for $p_m < 0$ than for $p_m > 0$, leading to a much larger 
asymmetry in this region as we would see in Fig.~\ref{fig2}. We also note that 
differences between complete RDWIA reduced cross section and $\rho_{DW}$ 
(hence deviations from factorization) are more noticeable at 
$|p_m| > 250$ MeV/c in the $p_m < 0$ region. Nonrelativistic calculations 
would generally yield results on the line of the EMA ones presented here. 
Note also that, the reduced cross 
section in EMA practically coincides with $\rho_{DW}$ for the $s_{1/2}$ 
orbital in $^{40}Ca$, and even in the $^{16}O$ $p_{1/2}$ and $p_{3/2}$ 
orbitals the reduced cross sections in EMA and $\rho_{DW}$ are rather close 
in the whole $p_m$ range. 

In Figs.~\ref{fig2} to~\ref{fig4} we show the $TL$ asymmetry, 
electron analyzing
power, induced polarization and transferred polarizations, respectively, 
for proton knockout from the $p_{1/2}$ (left panels), $p_{3/2}$ (middle) in 
$^{16}O$ and $s_{1/2}$ (right) shells in $^{40}Ca$. Results are computed for 
CC2 current operator and Coulomb gauge. The bound nucleon wave function 
corresponds to the set NLSH~\cite{Serot79,Hor81,Hor91,Sharma93} and the 
outgoing 
nucleon wave function has been derived using the EDAIO relativistic optical 
potential parameterization~\cite{Coo93}. As in the previous figure, 
the selected kinematics corresponds 
to the experimental conditions of the experiments E89003 and E89033 performed 
at Jlab~\cite{E89003,E89033,Gao}. This is $(q,\omega)$ constant kinematics 
with $q=1$~GeV/c, $\omega=440$ MeV and the electron beam energy fixed to 
$\varepsilon_i=2.445$ GeV. Coplanar kinematics, with $\phi_F = 0^{\circ}$, 
are chosen for computing the polarization asymmetries. Therefore, as
$P_l = P_s = P'_n = 0$ when $\phi_F = 0^{\circ}$, they are not plotted. 
In each graph, we show five curves corresponding to 
the following approaches: RDWIA (solid), RDWIA but without spin-orbit coupling 
in the final nucleon state, denoted as RDWIA-noLS (dashed), 
EMA (short-dashed), EMA-noLS (dotted), and finally the {\sl factorized} result 
(dash-dotted).

As shown in Sec.~III, factorization only holds within the EMA 
approach and assuming specific conditions on the spin dependence in the 
problem. In Table~1 we summarize the basic assumptions
within EMA that lead to factorization for the different observables. 
To simplify the discussion of the results that follows
we consider each observable separately. 

The asymmetry $A_{TL}$, presented in Fig.~\ref{fig2}, shows 
that factorization emerges
within EMA in the case of the $s_{1/2}$ shell (where EMA, EMA-noLS and 
{\sl factorized} results coincide). For spin-orbit dependent bound 
states ($p_{1/2}$ and $p_{3/2}$), factorization emerges only when there is no 
spin-orbit coupling in the final state (EMA-noLS coincides with 
{\sl factorized} results). Also note that the 
oscillatory behaviour shown by $A_{TL}$ in RDWIA and in RDWIA-noLS
is almost entirely lost within EMA, even when there is no factorization.
This reflects the crucial role played by the dynamical enhancement of the 
lower components of the nucleon wave functions for this observable. 
The spin dependence in the final nucleon state modifies significantly the 
values of $A_{TL}$ even at low missing momentum, but preserves its general 
oscillatory structure, compare for instance RDWIA vs RDWIA-noLS or EMA vs 
EMA-noLS.  

The electron analyzing power $A$ is presented in Fig.~\ref{fig3}. 
This observable is 
zero in coplanar kinematics so the azimuthal angle is fixed to 
$\phi_F=225^{\circ}$ in  Fig.~\ref{fig3}, but the remarks that follow also 
apply to other $\phi_F \ne 0^{\circ},180^{\circ}$ values. As we demonstrated 
in Sec.~III, the fifth 
response $R^{TL'}$ involved in $A$ only factorizes if there is no spin-orbit 
contribution in the initial and final nucleon wave functions. Moreover, in such
situation $R^{TL'}=0$ and hence $A=0$, as occurs for $s_{1/2}$ shell within 
EMA-noLS 
in Fig.~\ref{fig3}. From a careful inspection of 
Fig.~\ref{fig3} we also observe that the
main differences between the various approaches come from the spin-orbit term 
in the final state. Note that the discrepancy between RDWIA and EMA 
(or likewise between RDWIA-noLS and EMA-noLS) is significantly smaller than 
the discrepancy between RDWIA and RDWIA-noLS (or EMA vs EMA-noLS). In all of 
the cases with $A \neq 0$, oscillations survive. The behaviour of $A$ 
contrasts with the one observed for the asymmetry $A_{TL}$. This is due to 
the fact that factorization is broken down already at the EMA level even in 
the $s_{1/2}$ shell. 

The induced polarization $P_n$ is presented in Fig.~\ref{fig4}. 
Here the discussion of 
results follows similar trends to the previous one on $A$. Factorization 
requires no spin 
dependence in any of the nucleon wave functions, being the induced polarized 
responses equal to zero in such a case (notice that $P_n$ is zero in 
the plane wave limit). In any other situation factorization breaks down and 
$P_n$ shows strong oscillations in all cases. Again, it is important 
to point out that the behaviour of the RDWIA calculation is qualitatively 
followed by the EMA approach, differing much more from the RDWIA-noLS or 
EMA-noLS. This reveals the important effects introduced by the spin-orbit 
coupling in the optical potential for polarized observables, contrary to what 
happens for the unpolarized $A_{TL}$. 

The comment above applies also to the transferred polarizations $P'_l$ and 
$P'_s$ (Fig.~\ref{fig5}) for which RDWIA and EMA 
approaches give rise to rather similar 
oscillating (unfactorized) results. On the contrary, RDWIA-noLS, which is also 
unfactorized, deviates  significantly from RDWIA due to the crucial role of 
the spin-orbit dependence in the final state. Finally, EMA-noLS coincides with 
the bare asymmetries showing a flat behaviour without oscillations.
This is in accord with the findings in Sec.~III~B, where we 
demonstrated that the unpolarized $R^\alpha$ and transferred polarized 
$R_{l,s}^{\alpha'}$ responses factorize with the same momentum distribution 
function (see Table~1 and Eq.~(\ref{wmununolsf2})).

\section{Summary and conclusions}

A systematic study of the property of factorization in quasielastic 
$(\vec{e},e'\vec{p})$ reactions has been presented. Starting from a RDWIA 
analysis, we have reformulated the EMA approach and studied the conditions 
which are needed to get factorization. In this context, we have explored the 
role of the spin-orbit coupling in the initial and/or final nucleon states 
and its influence on the breakdown of factorization.

From our general study we conclude that exact factorization only emerges 
within the EMA approach, i.e., neglecting the dynamical enhancement of the 
lower components in the nucleon wave functions by using Eq.~(\ref{asymp}). 
Furthermore, additional 
restrictions on the spin dependence in the problem are necessary to get 
factorization in the case of polarized observables.

Within the EMA approach, the factorization properties of various 
$(\vec{e},e'\vec{p}\,)$ responses and asymmetries are as follows (see also 
Table~I):
 
The unpolarized $R^{\alpha}$ responses factorize to a single, polarization
independent, momentum distribution when the initial {\sl or} the final nucleon 
wave
functions are independent on spin-orbit coupling (i.e., depend on $\ell$ but
not on $j$). As a consequence, the $A_{TL}$ asymmetry is in these cases
given by the bare-nucleon $A_{TL}^{bare}$ asymmetry.
 
The fifth response $R^{TL'}$ (and consequently $A$), depending on electron
beam polarization, never factorizes, but becomes zero when both initial and 
final nucleon wave functions are independent on spin-orbit coupling, as well 
as in the nonrelativistic plane wave limit (PWIA).

The transferred polarization responses $R^{\alpha '}_k$ factorize when the 
final nucleon wave function is independent on $j$. Consequently the 
transferred polarizations are in this case given by the bare-nucleon ones, 
independent on whether the initial state may or may not depend on spin-orbit 
coupling.
 
The induced polarized responses $R^{\alpha}_k$ do not factorize even when the
final nucleon wave function is independent on $j$, unless the initial wave
function is also independent on $j$, in which case $R^{\alpha}_k$ 
become zero. If the final wave
function depends on spin-orbit coupling but the initial wave function does not,
the induced polarized responses factorize with a polarization-dependent
momentum distribution different from the unpolarized one. Therefore, as stated
in Eq.~(\ref{pp}), a new factorization property emerges when there is no
spin-orbit coupling in the initial state.

From our numerical calculations a clear difference in the behaviour 
of polarized and unpolarized observables comes out. In the case of the 
unpolarized $A_{TL}$ asymmetry, its general structure is not substantially 
modified  by the final spin-orbit dependence, being much more affected by the 
lower components of the nucleon states. The {\sl strong} oscillations in 
$A_{TL}$ within RDWIA practically vanish in EMA. On the contrary, the 
polarized asymmetries $A$, $P_n$ and $P'_{l,s}$, present a very strong 
sensitivity to the final spin dependence, while the general structure of the 
RDWIA results is preserved by the EMA calculations.

As a general conclusion, we can say that observables that require less extra 
assumptions (apart from EMA) to factorize, are more sensitive to any 
ingredient of the calculation that breaks factorization. Such observables are 
good candidates to test the elements of any model/calculation, 
as it is the case of the $A_{TL}$ asymmetry.

In spite of the fact that factorization is not reached when realistic 
calculations are made, we show that the reduced cross sections extracted 
from fully unfactorized calculations  
follow the factorized distorted momentum distribution quite well for moderate 
values of the missing momentum, where the bulk of the cross section lies. Then, 
reduced cross sections and integrated quantities directly related to them, 
like nuclear transparencies or inclusive cross sections, are reasonably 
predicted by the factorized scheme, as long as one remains at quasielastic 
kinematics. We may conclude that the unpolarized cross section follows 
closely the factorized calculation that takes FSI into account. In other 
words, in spite of the breakdown of factorization of the cross section 
introduced by FSI and by negative energy components of the relativistic model, 
one may still extract a meaningful effective momentum distribution within this 
formalism.

While the bulk of the cross section factorizes to a good approximation, ratios 
of cross sections like $A_{TL}$ or polarizations are very sensitive to the 
ingredients of the calculation that break factorization. This is why in 
particular the $A_{TL}$ observable is very sensitive to the negative energy 
components of the wave functions, and provides a plausible signature of the 
relativistic dynamics.  

Contrary to $A_{TL}$, polarizations are much more sensitive to the spin-orbit 
properties of the upper components of the wave functions than to the dynamical 
enhancement of the lower components. Yet, RDWIA transferred polarizations 
closely match the factorized results in certain $p_m$ ranges. This suggests 
that measuring transferred polarizations in those ranges may safely explore 
modifications of the nucleon form factor ratios in the nuclear medium. 

\subsection*{Acknowledgements}

The authors thank T.W. Donnelly for his helpful comments.
This work was partially supported by funds provided by DGI (Spain) 
under Contracts Nos. BFM2002-03315, BFM2002-03562, 
FPA2002-04181-C04-04 and BFM2003-04147-C02-01 and by the Junta de 
Andaluc\'{\i}a (Spain). J.R.V. and  M.C.M. acknowledge financial support 
from the Consejer\'{\i}a de Educaci\'on de la Comunidad de Madrid and the 
Fundaci\'on C\'amara (University of Sevilla), respectively.

\section*{Appendix A}

In this appendix we present in more detail the hadronic bare-nucleon tensor 
(\ref{calwmunu}), which can be written using traces in the form
\beqa
{\cal W}^{\mu\nu}_{ss',\, hh'}=Tr\left[ \overline{\widehat{J}}\,^\mu 
u(\np_F,s) \overline{u}(\np_F,s')\widehat{J}^\nu u(\np_I,h') \overline{u}(\np_I,h) 
\right] \, ,
\eeqa
where we use the notation 
$\overline{\widehat{J}}^\mu\equiv \gamma^0\widehat{J}^{\mu \dagger} \gamma^0$.

Making use of the following relation~\cite{Cab93,Cab98}
\beqa
u(\np,s) \overline{u}(\np,s')=\frac{\delta_{ss'} +\gamma_5 \phibar_{ss'}}{2} \, \frac{\Pbar+M}{2M} \, ,
\eeqa
with $\varphi_{ss'}^\mu$ a pseudovector defined as
$\varphi_{ss'}^\mu=\overline{u}(\np,s') \gamma^\mu\gamma^5 u(\np,s)$ which 
reduces to the four spin $S^\mu$ in the diagonal case, i.e., 
$\varphi_{ss}^\mu = S^\mu$, the bare nucleon tensor reads
\beqa
{\cal W}^{\mu\nu}_{ss',\, hh'}&=&
\frac{1}{16M^2}
Tr\left[\overline{\widehat{J}}\,^\mu (\Pbar_F+M) \widehat{J}^\nu (\Pbar_I+M) \right]
\, \delta_{ss'} \, \delta_{hh'} 
\nonumber \\
&+&
\frac{1}{16M^2} Tr\left[\overline{\widehat{J}}\,^\mu (\Pbar_F+M) \widehat{J}^\nu
\gamma_5 \phibar_{hh'} (\Pbar_I+M) \right] \, \delta_{ss'}
\nonumber \\
&+& 
\frac{1}{16M^2} Tr\left[\overline{\widehat{J}}\,^\mu \gamma_5 \phibar_{ss'}
(\Pbar_F+M) \widehat{J}^\nu (\Pbar_I+M) \right] \, \delta_{hh'}
\nonumber \\
&+&
\frac{1}{16M^2} Tr\left[\overline{\widehat{J}}\,^\mu \gamma_5 \phibar_{ss'} 
(\Pbar_F+M) \widehat{J}^\nu
\gamma_5 \phibar_{hh'} (\Pbar_I+M) \right] \, .
\label{app1}
\eeqa
This result is expressed in a compact form in Eq.~(\ref{calwmunu2}).

\section*{Appendix B}

Let us consider the case of no spin-orbit coupling in the final nucleon wave
function. This means that the radial functions 
$g_{\kappa}$ and $\delta_{\kappa}$ depend only on 
$l$ but not on $j$. Then the upper component of the wave function is given by
\beqa 
\psi_{F, \, up}^{s_F}(\np) &=& 
4\pi \sqrt{\frac{E_F+M}{2E_F}}\sum_{\ell m} e^{-i\delta^\ast_\ell}
Y_\ell ^{m \ast}(\widehat{\np}_F)g^\ast_l(p)
\sum_{j \mu} \langle \ell m \frac{1}{2} s_F |j \mu \rangle 
\Phi_{\kappa}^{\mu}(\widehat{\np})
\nonumber \\
&= &
G(\np,\np_F)\chi_{s_F}
\eeqa
with
\beqa
G(\np,\np_F) &=&
       4\pi \sqrt{\frac{E_F+M}{2E_F}}\sum_{\ell m} e^{-i\delta^\ast_\ell}
Y_\ell ^{m \ast}(\widehat{\np}_F)g^\ast_l(p)Y_\ell ^m(\widehat{\np}) \, .
\label{functiong}
\eeqa
The resulting wave function for the ejected proton is then
\beqa
& &\psi_F^{s_F \, EMA}(\np)
=\sqrt{\frac{2M}{E_F+M}} G(\np,\np_F)
u(\np_F,s_F) \, .
\eeqa
Introducing this result into the expression of the current matrix 
element, we get
\beqa
J^\mu_{EMA} &=&
\sum_{m_{\ell_b} h}
\langle \ell_b m_{\ell_b} \frac{1}{2} h |j_b \mu_b \rangle
\left[\overline{u}(\np_F, s_F)\widehat{J}^\mu u(\np_I, h)\right]
U^{m_{\ell_b}}_{\kappa_b}(\np_F,\nq)
\label{jemanolsf}
\eeqa
being,
\beqa
U^{m_{\ell_b}}_{\kappa_b}(\np_F,\nq)=
\frac{2M}{\sqrt{(E_I+M)(E_F+M)}} (-i)^{\ell_b}
\int d\np \, G^\ast(\np+\nq,\np_F)
g_{\kappa_b}(p)
Y_{\ell_b}^{m_{\ell_b}}(\widehat{\np}) \, .
\label{utilde}
\eeqa

We observe that the whole dependence on the spin polarization $s_F$ is contained in the 
Dirac spinor $u(\np_F, s_F)$.
From Eq.~(\ref{jemanolsf}) we can immediately construct the hadronic tensor 
$W^{\mu\nu}_{EMA}$, which can be written in the form of Eq.~(\ref{wmununolsf}) 
with the momentum distribution function given by
\beq
\widetilde N_{hh'}(\np_F,\nq) = \frac{1}{2j_b+1} \sum_{\mu_b}
\sum_{m_{\ell_b}m_{\ell_b}'}
\langle \ell_b m_{\ell_b} \frac{1}{2} h |j_b \mu_b \rangle
\langle \ell_b m_{\ell_b}' \frac{1}{2} h' |j_b \mu_b \rangle
U^{m_{\ell_b} \ast}_{\kappa_b} U^{m_{\ell_b}'}_{\kappa_b} \, .
\label{ntilde}
\eeq

As a particular example,
let us consider the case of the plane wave limit without dynamical relativistic
effects. In this situation the function $G(\np,\np_F)$ (Eq.~(\ref{functiong})) 
simply reduces to
\be
G^{PW}(\np,\np_F)=\sqrt{\frac{E_F+M}{2E_F}}(2\pi)^{3/2}\delta^3(\np-\np_F)
\ee
and the momentum distribution results
\beqa
\widetilde N_{hh'}^{PW}(\np_F,\nq) &=&
\delta_{hh'} \, \frac{M^2}{2 E_I E_F}(2\pi)^3 \,  N^{EMA}(p_I) 
\, .
\label{ntildpw}
\eeqa


\begin{table}[h]
\begin{center}
\begin{tabular}{|c|c|c|c|}
  & no LS initial   & no LS final     & no LS both \\  \hline
$A_{TL}$   & $A_{TL}^{bare}$ & $A_{TL}^{bare}$ & $A_{TL}^{bare}$ \\  \hline
$A$  &    ---          &  ---            &    0             \\  \hline
$P_n$      &    ---         & ---            &    0             \\  \hline
$P_l$      &    ---         & ---            &    0             \\  \hline
$P_s$      &    ---         & ---            &    0             \\  \hline
$P_n'$      &    ---         &  $P_n^{'bare}$   & $P_n^{'bare}$       \\ \hline
$P_l'$      &    ---         &  $P_l^{'bare}$   & $P_l^{'bare}$       \\ \hline
$P_s'$      &    ---         &  $P_s^{'bare} $   & $P_s^{'bare}$       \\  
\end{tabular}
\vspace{0.4cm}
\caption[Table 1]{Properties of factorization of different observables
using the EMA approximation and turning off the spin-orbit coupling in
the initial wave function (first column), in the final wave function (second
column) or in both simultaneously (third column).}
\end{center}
\end{table}


\begin{figure}
{\par\centering  \resizebox*{0.5\textheight}{0.9\textwidth}{\rotatebox{0}
{\includegraphics{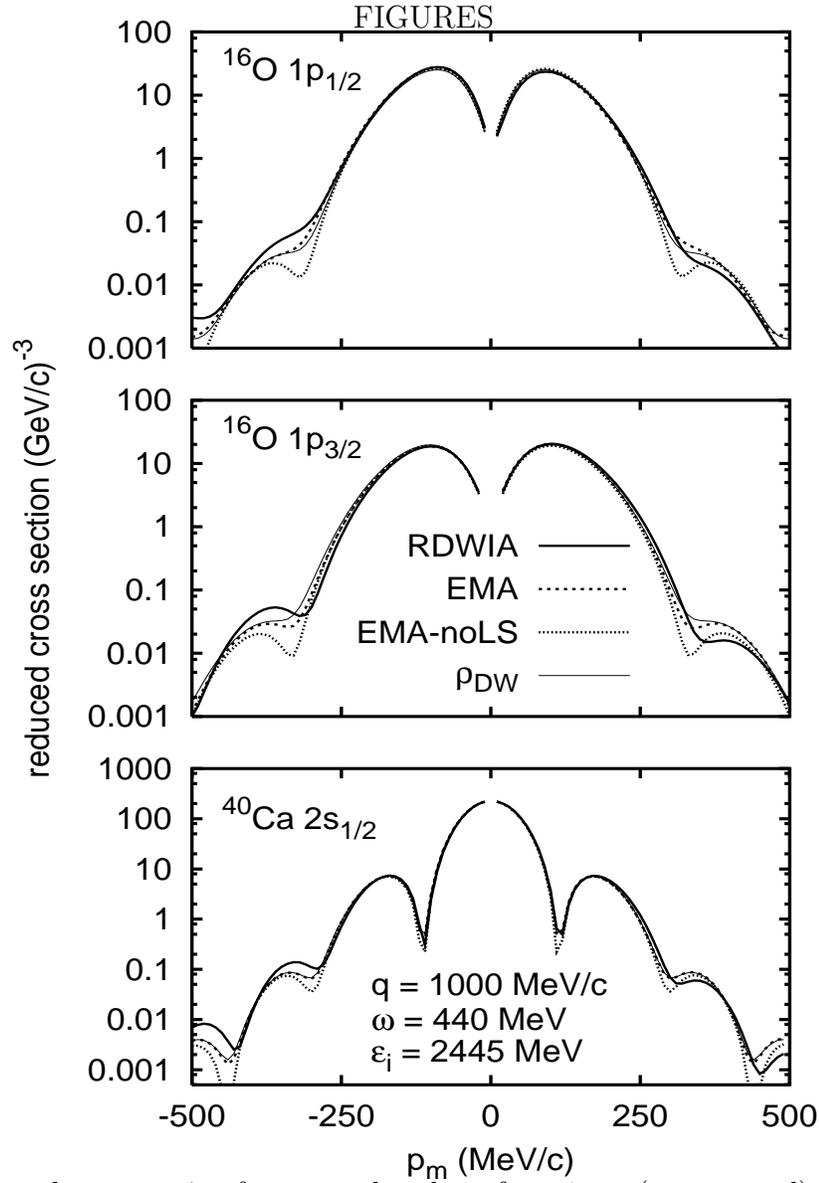}}} \par}
\caption{Reduced cross section for proton knockout from
\protect\( 1p_{1/2}\protect \) (upper panel) and
\protect\( 1p_{3/2}\protect \) (middle panel) in \protect\( ^{16}O\protect \) 
and from \protect\( 2s_{1/2}\protect \) in
\protect\( ^{40}Ca\protect \) (lower panel). 
RDWIA calculations (solid line) are compared to EMA
(short-dashed line) and EMA-noLS (dotted line) results. The corresponding
relativistic distorted wave momentum distribution is also plotted 
(thin solid line). Negative (positive) $p_m$ values correspond to 
$\phi_F = 0^{\circ}$ ($180^{\circ}$) respectively.
\label{fig1}}
\end{figure}
 
\begin{figure}
{\par\centering  \resizebox*{0.6\textheight}{0.5\textwidth}{\rotatebox{-90}
{\includegraphics{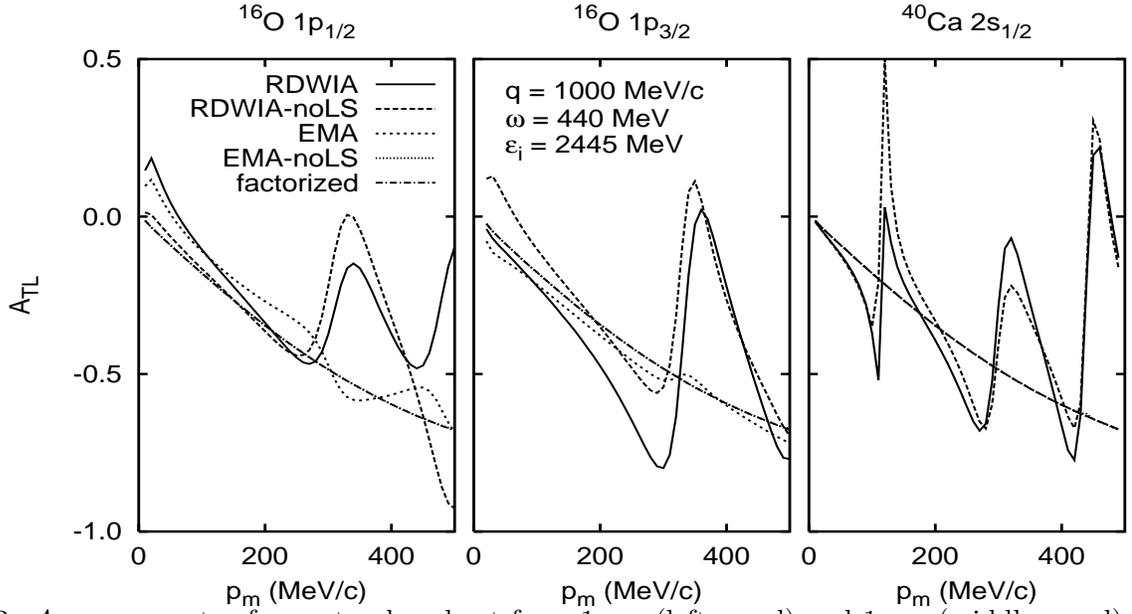}}} \par}
\caption{$A_{TL}$ asymmetry for proton knockout from 
\protect\( 1p_{1/2}\protect \) (left panel) and \protect\( 1p_{3/2}\protect \) 
(middle panel) in \protect\( ^{16}O\protect \) and from 
\protect\( 2s_{1/2}\protect \) in \protect\( ^{40}Ca\protect \) (right panel).
RDWIA calculations (solid line) are compared to RDWIA-noLS
(dashed line), EMA (short-dashed line), EMA-noLS (dotted line) and
factorized (dash-dotted line) results. The EMA-noLS calculation coincides in
all panels with the factorized ($A_{TL}^{bare}$) result. In the right hand
panel EMA, as well as EMA-noLS, coincides with the factorized
($A_{TL}^{bare}$) result.
\label{fig2}}
\end{figure}

\begin{figure}
{\par\centering  \resizebox*{0.6\textheight}{0.5\textwidth}{\rotatebox{-90}
{\includegraphics{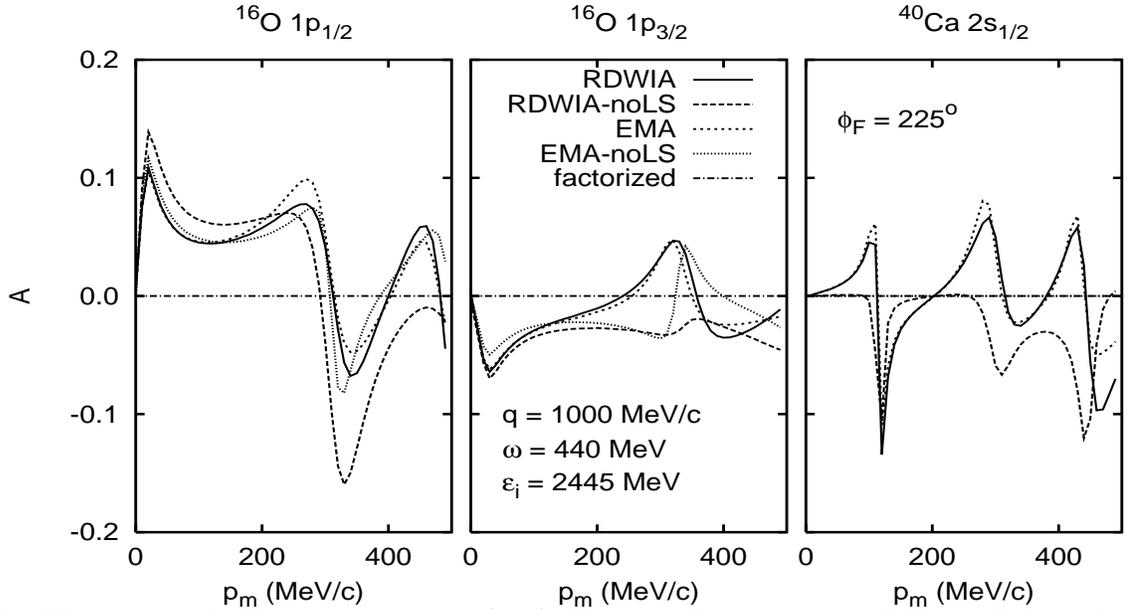}}} \par}
\caption{Electron analyzing power $A$ at $(q,\omega)$ constant kinematics and 
azimuthal angle $\phi_F=225^{\circ}$. The labeling of the curves is as in
Fig.~\ref{fig2}. For this observable factorization is only achieved in the 
EMA-noLS curve on the right hand panel. See text for details. 
\label{fig3}}
\end{figure}

\begin{figure}
\begin{center}
{\par\centering \resizebox*{0.6\textheight}{0.5\textwidth}{\rotatebox{-90}
{\includegraphics{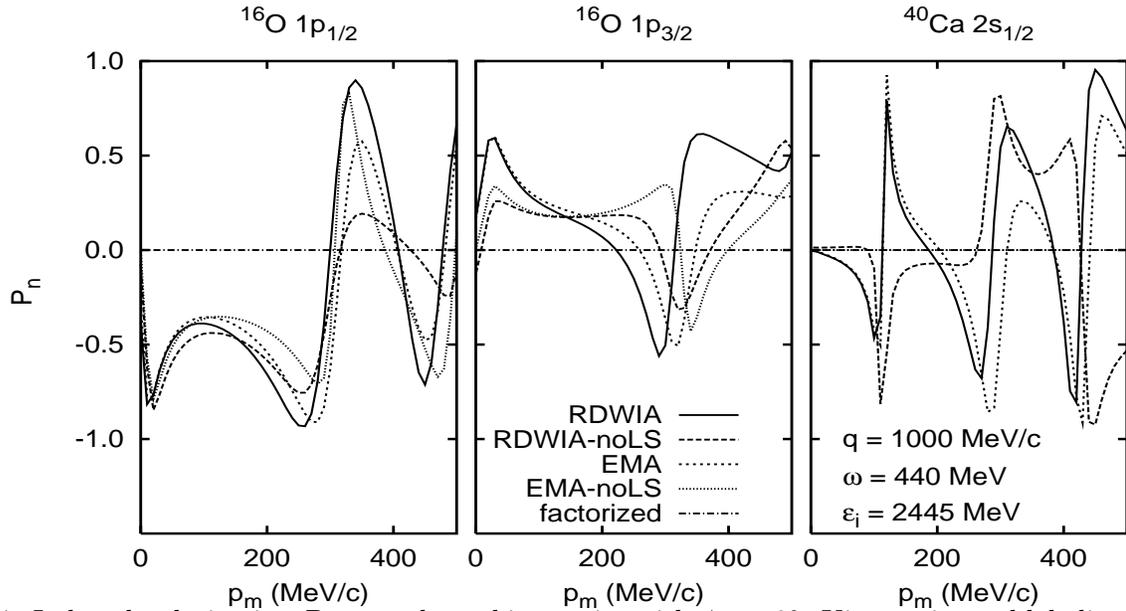}}} \par}
\caption{ Induced polarization $P_n$ at coplanar kinematics with 
$\phi_F = 0^{\circ}$. Kinematics and labeling as in 
Fig.~\ref{fig2}. Only the EMA-noLS 
calculation for a $s_{1/2}$ shell factorizes.
\label{fig4}}
\end{center}
\end{figure}

\begin{figure}
{\par\centering  \resizebox*{0.6\textheight}{0.5\textwidth}{\rotatebox{-90}
{\includegraphics{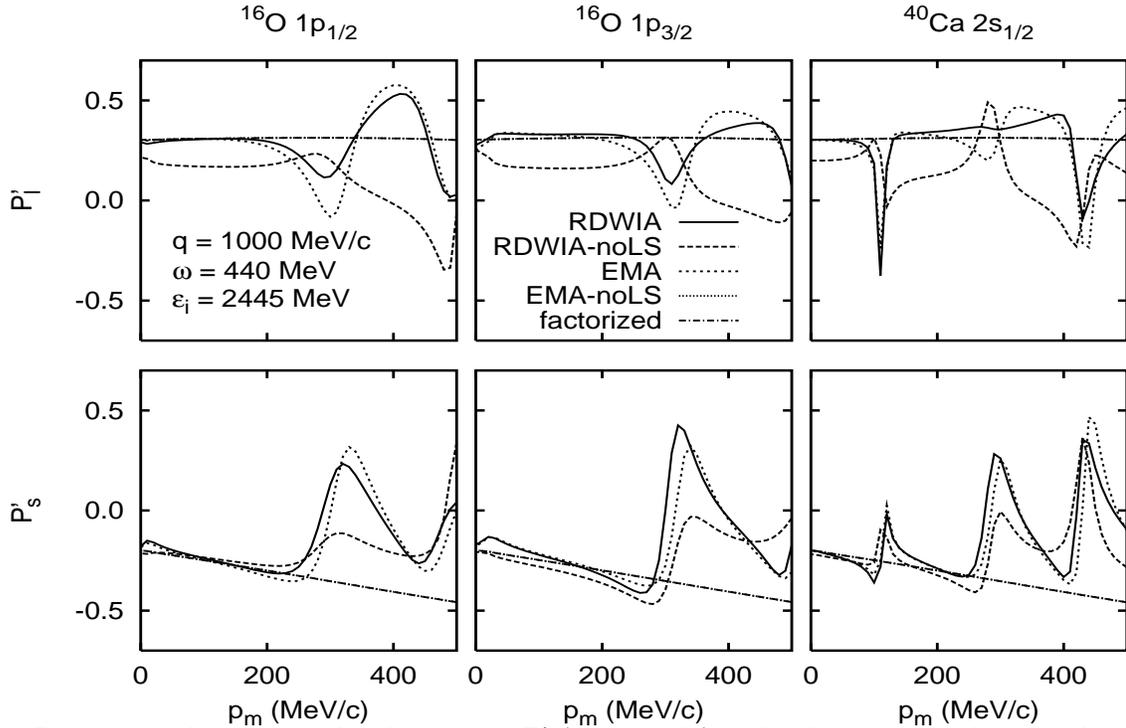}}} \par}
\caption{ Longitudinal transferred polarization $P'_l$ (top panels) and 
sideways transferred polarization $P'_s$ (bottom panels) at coplanar 
kinematics ($\phi_F = 0^{\circ}$). In this case, 
factorization is obtained within the EMA approach when there is no spin-orbit 
coupling in the final state (EMA-noLS, dotted line). 
\label{fig5}}
\end{figure}


\end{document}